\documentclass[12pt]{article}
\usepackage{amsmath,amsfonts}
\usepackage{graphicx}   
\usepackage[margin=3cm]{geometry}    
\newcommand{\Cx}{{\mathbb C}}
\newcommand{\Ir}{{\mathbb Z}}
\newcommand{\Rl}{{\mathbb R}}
\newcommand{\A}{{\mathcal A}}
\newcommand{\B}{{\mathcal B}}
\renewcommand{\H}{{\mathcal H}}

\def\idty{{\mathchoice {\mathrm{1\mskip-4mu l}} {\mathrm{1\mskip-4mu l}} %
{\mathrm{1\mskip-4.5mu l}} {\mathrm{1\mskip-5mu l}}}}

\newcommand{\spec}{\mathop{\rm spec}}
\newcommand{\dom}{\mathop{\rm Dom}}

\renewcommand{\vec}[1]{\boldsymbol{#1}}

\newcommand{\ket}[1]{\left\vert #1\right\rangle}

\newcommand{\be}{\begin{equation}}
\newcommand{\ee}{\end{equation}}
\newcommand{\bea}{\begin{eqnarray}}
\newcommand{\eea}{\end{eqnarray}}
\newcommand{\beann}{\begin{eqnarray*}}
\newcommand{\eeann}{\end{eqnarray*}}
\newcommand{\eq}[1]{(\ref{#1})}
\newtheorem{theorem}{Theorem}[section]

\newtheorem{conjecture}[theorem]{Conjecture}
\begin{document}
\renewcommand{\thefootnote}{\fnsymbol{footnote}}
\title{Recent Progress in Quantum Spin Systems$^*$}
\author{Bruno Nachtergaele and Robert Sims\\[10pt]
Department of Mathematics\\ University of California at Davis\\
Davis CA 95616, USA\\
Email: bxn@math.ucdavis.edu, rjsims@math.ucdavis.edu}
\date{Version: \today }
\maketitle
\bigskip\begin{center}
{\em This paper is dedicated to the memory of John T. Lewis.}
\end{center}

\abstract{Some recent developments in the theory of quantum spin systems 
are reviewed.}

\footnotetext[1]{Copyright \copyright\ 2005 by the authors. 
This paper may be reproduced, in its entirety, for non-commercial purposes.}

\section{Introduction}
\label{sec:intro}

We will review recent results on quantum systems grouped into four sections: 
Decay of Correlations (Section \ref{sec:decay}), Perturbation 
Theory (Section \ref{sec:perturb}), Ferromagnetic Ordering of Energy Levels 
(Section \ref{sec:foel}), and Droplet Excitations of the XXZ Chain 
(Section \ref{sec:droplets}).

A quantum spin is any quantum system with a finite-dimensional
Hilbert space of states. A quantum spin system, defined over a set $V$,
consists of a finite or infinite number of spins, each of which labeled by 
some $x \in V$. When $V$ is an infinite set, typically corresponding to
the vertices of a lattice or a graph, one often considers families of 
quantum spin systems, labeled by the finite subsets $X \subset
V$. Certain properties are easily stated for infinite sets $V$
directly, but, for now, we will assume that $V$ is finite. In this case, 
the Hilbert space
of states is
$$
\H_V=\bigotimes_{x\in V} \Cx^{n_x},
$$
where the dimensions $n_x\geq 2$ are related to the magnitude of the spins, $s_x\in\{1/2,1,3/2,\ldots\}$,
by $n_x=2s_x+1$.
For each spin, the basic observables are the complex $n \times n$
matrices, which we will denote by $M_n$. The algebra of 
observables for the system is then
$$
\A_V=\bigotimes_{x\in V} M_{n_x}=\B(\H_V).
$$
Given a Hamiltonian, a self-adjoint observable $H_V = H_V^* \in \A_V$,
one may generate the Heisenberg dynamics of the system for any $A\in
\A_V$ and $t \in \Rl$ by setting 
$$
\tau^V_t(A)=U^*_V(t) A U_V(t)
$$
where $U_V(t)=e^{-itH_V}$.

One of the most important examples of a quantum spin system is the Heisenberg
model. In general, this model is defined on a graph $(V,E)$ which consists of a
set of vertices $V$ and an edge set $E$ comprised of pairs of vertices denoted
by $e=(xy)$ for $x,y\in V$. Let $S^i_x$, $i=1,2,3$, denote the standard spin
$s_x$ matrices associated with the vertex $x$, and for each edge $e=(xy)$, let
$J_{xy} \in \Rl$ be a coupling constant corresponding to $e$. The Heisenberg
Hamiltonian (also called the XXX Hamiltonian) is then given by
\be
H_V=-\sum_{(xy)\in E} J_{xy} \vec{S}_x\cdot \vec{S}_y \, ,
\label{hh}\ee
where $\vec{S}_x$ denotes the vector with components $S_x^1,S_x^2,S_x^3$.
The most commonly studied models are those defined on a lattice, such
as $\Ir^d$, with translation invariance. For such Hamiltonians, the
magnitude of the spins is constant, i.e., $s_x=s$, the edges
are the pairs $(xy)$ such that $\vert x-y\vert=1$, and  $J_{xy}
=J$. Depending of the sign of $J$, the Heisenberg model is said to 
be ferromagnetic ($J>0$) or the antiferromagnetic ($J<0$).

For extended systems, i.e., those corresponding to sets $V$ of infinite
cardinality, more care is needed in defining the quantities 
mentioned above. Hamiltonians are introduced as a sum of local terms
described by an interaction, a map $\Phi$ from the set of 
finite subsets of $V$ to $\A_V$, with the property that 
for each finite $X \subset V$, $\Phi(X) \in \A_X$ and $\Phi(X) = \Phi(X)^*$.
Given an interaction $\Phi$, the Hamiltonian is defined by
$$
H_V=\sum_{X\subset V} \Phi(X).
$$
Such infinite systems are often analyzed by considering families of 
finite systems, indexed by the subsets of $V$, and taking the
appropriate limits. For example, the $C^*$-algebra of 
observables, $\A$, is defined to be the norm completion of the 
union of the local observable algebras $\bigcup_{X\subset V}\A_X$.

Since we want to discuss decay of spatial correlations, we need a
distance function on $V$. Let $V$ be equipped with a metric $d$. 
For typical examples, $V$ will be a graph and $d$ will
be chosen as the graph distance: $d(x,y)$ is the 
length of the shortest path (least number of edges)
connecting $x$ and $y$. The diameter, $D(X)$, of a finite subset 
$X \subset V$ is
$$
D(X)=\max \{ d(x,y)\mid x,y\in X\}.
$$

In order for the finite-volume dynamics to converge to a strongly continuous
one-parameter group of automorphisms on $\A$, one needs to impose
a decay condition on the interaction. For the sake of brevity, we will
merely introduce the norm on the interactions that will later appear
in the statement of our results. For weaker conditions which ensure
existence of the dynamics see \cite{bratteli1997,simon1993,matsui1998}.
We will assume that the dimensions $n_x$ are bounded:
$$
N=\sup_{x\in V} n_x < \infty ,
$$
and that there exists a $\lambda >0$ such that the following
quantity is finite:
$$
\Vert \Phi \Vert_{\lambda} \, := \, \sup_{x \in V} \, \sum_{X \ni x}
\, |X|  \, \Vert \Phi(X) \Vert\, N^{2|X|} \, e^{ \lambda D(X)} < \infty.
$$
Under these conditions, one can prove quasi-locality of the dynamics,
in the sense that, up to exponentially small corrections, there is 
a finite speed of propagation. 
In the next section, we give a precise statement
of this result and apply it to prove that a nonvanishing spectral gap above the 
ground state energy implies exponential decay of spatial correlations in the 
ground state. This can be regarded as a non-relativistic analogue
of the
Exponential Clustering Theorem in relativistic quantum field theory 
\cite{fredenhagen1985}.
The idea that a Lieb-Robinson bound can be used as a replacement
for strict locality in the relativistic context 
can be found in \cite{hastings2004}.

\section{Decay of Correlations}\label{sec:decay}

\subsection{Lieb-Robinson Bounds}

Our proof of exponential clustering uses a generalization of the well-known
theorem by Lieb and Robinson \cite{lieb1972}.
The aim of such a result is to prove quasi-locality of the dynamics,
expressed as an estimate for commutators of the form
$$
\left[ \, \tau_t(A) \, , \, B \, \right],
$$
where $t\in\Rl$, $A \in \A_X$, $B \in \A_Y$, and $X,Y\subset V$.
Clearly, such commutators vanish if $t=0$ and $X\cap Y=\emptyset$.
Quasi-locality, or finite group-velocity, as the property is also called,
means that the commutator remains small up to a time proportional
to the distance between $X$ and $Y$.

It will be useful to consider the following quantity
$$
C_B(x,t) :=  \sup_{A \in\A_{x}} \frac{ \| \left[ \, \tau_t(A) \, , \,
    B \, \right] \|}{ \| A \|} \, ,
$$
for $x\in V$, $t\in \Rl$, $B\in \A_V$.
The basic result obtained by Lieb and Robinson in the case
of translation invariant systems on a lattice, and by us in the present
setup, is the following theorem \cite{nachtergaele2005d}.

\begin{theorem}\label{thm:lr}
For $x\in V$, $t\in \Rl$, and $B\in\A_V$, we have the bound
\beann
C_B(x,t) &\leq& e^{2\, |t| \, \| \Phi \|_{\lambda}} C_B(x,0)\\
&&+  \sum_{y \in V: y \neq x} \,e^{- \, \lambda \,
  d(x,y)}  \left( e^{2 \, |t| \, \| \Phi \|_{\lambda}} - 1
\right)C_B(y,0) \, .
\eeann
\end{theorem}

Our proof avoids the use of the Fourier transform which seemed essential
in the work by Lieb and Robinson and appeared to be the main obstacle
to generalize the result to non-lattice $(V,d)$.

If the supports of $A$ and $B$ overlap, then the trivial bound $ \Vert
[\tau_t(A),B] \Vert \leq 2 \Vert A \Vert \Vert B \Vert$ is
better. Observe that for $B \in \A_Y$, one has that
 $C_B(y,0)\leq 2\Vert B \Vert \, \chi_Y(y)$, where
$\chi_Y$ is the characteristic function of $Y$, and therefore if $x
\not\in Y$, then one obtains for any $A \in \A_x$ a bound of the form
$$
\Vert [\tau_t(A),B]\Vert\leq 2\vert Y\vert \, \Vert A\Vert \Vert B\Vert 
\left(e^{2 \, |t| \, \| \Phi \|_{\lambda}} -1\right) e^{- \, \lambda
  \, d(x,Y)} \, , 
$$
where
$$
d(x,Y)=\min \{ d(x,y)\mid y\in Y\} \, .
$$
Moreover, for general local observables $A\in\A_X$, one may estmate
$$
\Vert [\tau_t(A),B]\Vert\leq N^{2\vert X\vert}\Vert A\Vert\sum_{x\in X} 
C_B(x,t) \, ,
$$
in which case, Theorem~\ref{thm:lr} provides a related bound.

\subsection{Exponential Clustering}

In the physics literature the term {\em massive ground state} implies two
properties: a spectral gap above the ground state energy and exponential decay
of spatial correlations. It has long been believed that the first implies the
second, and our next theorem proves that this is indeed the case. The converse,
that exponential decay must be necessarily accompanied by a gap is not true in
general. Exceptions to the latter have been known for some time, and it is not
hard to imagine that a  spectral gap can close without affecting the ground
state \cite{nachtergaele1996}.

For simplicity of the presentation, we will restrict ourselves to the case where we have 
a representation of the system (say, the GNS representation) in which the model
has a unique ground state. This includes most cases with a spontaneously
broken discrete symmetry. Specifically, we will assume that our system
is represented on a Hilbert space $\H$,
with a corresponding Hamiltonian $H \geq 0$, and that $\Omega \in \H$ is, up to a
phase, the unique normalized vector state for which $H \Omega =0$. 
We say that the system has a spectral gap if there exists $\delta >0$
such that $\spec (H) \cap (0,\delta) =\emptyset$, and in this case, 
the spectral gap, $\gamma$, is defined by
$$
\gamma=\sup\{\delta > 0 \mid \spec(H) \cap (0,\delta) =\emptyset\}.
$$

Our theorem on exponential clustering derives a bound 
for ground state correlations which take the form
\be
\langle\Omega, A\tau_{ib}(B)\Omega\rangle
\ee
where $b \geq 0$ and $A$ and $B$ are local observables. 
The case $b=0$ is the standard (equal-time) correlation
function. It is convenient to also assume a  
minimum site spacing among the vertices: 
\be
\inf_{ \stackrel{x,y \in V}{x \neq y}} d(x,y) =: a >0.
\ee
We proved the following theorem in \cite{nachtergaele2005d}.

\begin{theorem}[Exponential Clustering]\label{thm:decay}
There exists $\mu>0$ such that for any $x \neq y \in V$ and 
all $A\in\A_x$, $B\in\A_y$ for which
$\langle \Omega, B\Omega \rangle =0$, and $b$ sufficiently small,  
there is a constant $c(A,B)$ such that

\be
\left\vert \langle\Omega, A\tau_{ib}(B)\Omega\rangle \right\vert
\leq c(A,B)e^{-\mu d(x,y)\left( 1 + \frac{\gamma^2 b^2}{4\mu^2d(x,y)^2}\right)}
\, .
\label{decay}\ee
One can choose 
\be
\mu = \frac{\gamma\lambda}{4 \Vert \Phi\Vert_\lambda + \gamma}\, ,
\ee
and the bound is valid for $0\leq \gamma b\leq 2\mu d(x,y)$.
\end{theorem}

The constant $c(A,B)$, which can also be made explicit, depends 
only on the norms of $A$ and $B$, (in its more general form) 
the size of their supports, and the system's minimum vertex spacing
$a$. For $b=0$, Theorem~\ref{thm:decay} may be restated as
\be
\left\vert \langle\Omega, AB\Omega\rangle
-  \langle\Omega, A\Omega\rangle\, \langle\Omega, B\Omega\rangle\right\vert 
\leq c(A,B) e^{-\mu d(x,y)} .
\label{zerob}\ee
One may note that there is a trivial bound for large $b>0$
\be
\left\vert \langle\Omega, A\tau_{ib}(B)\Omega\rangle \right\vert
\leq \Vert A\Vert \, \Vert B\Vert \, e^{-\gamma b} \, .
\ee
In the small $b>0$ regime, the estimate (\ref{decay}) can be viewed as 
a perturbation of (\ref{zerob}). Often, the important observation is
that the decay estimate (\ref{decay}) is uniform in the imaginary time $ib$,
for $b$ in some interval whose length, however, depends on
$d(x,y)$. 

In a recent work, Hastings and Koma have obtained an analogous result 
for models with long range interactions \cite{hastings2005}.

\section{Perturbation Theory}\label{sec:perturb}

A major goal in the perturbation theory of quantum spin systems is to show that
the set of interactions for which the model has a unique ground state with a
non-vanishing spectral gap above it (in the thermodynamic limit), is open in a
suitable topology on the space of interactions. Significant steps toward this
goal have been made by a number of authors. Typically, the results obtained
apply to quantum perturbations of classical models in various degrees of
generality \cite{datta1996,borgs1996,borgs1997}. The remarkable paper by Kennedy
and Tasaki \cite{kennedy1992} was perhaps the first to make a serious attempt
to  get away from perturbing classical models. The new result by Yarotsky, which
we discuss here, can be seen as taking that line of approach one step further. 

Yarotsky's result makes it possible to prove stability of the massive phase 
provided that there is a nearby Generalized Valence Bond Solid model
\cite{affleck1987,fannes1992a,nachtergaele1996} that can be used as a reference 
point for the perturbation in the space of interactions. In particular, 
Yarotsky \cite{yarotsky2004} proves that the spin-1 chain with Hamiltonian 
\cite{affleck1987}
\be
H^{\rm AKLT}=\sum_x \left[\frac{1}{2}{\bf S}_x\cdot{\bf S}_{x+1}
+\frac{1}{6}({\bf S}_x\cdot{\bf S}_{x+1})^2+1/3\right]
\label{aklt}\ee
is contained in an open set of interactions with this property.

A very useful general theorem proved by Yarotsky can be stated for
the following class of models defined on $\Ir^d$, $d\geq 1$.
For these models the Hilbert space $\H_x$, at $x\in\Ir^d$, 
is allowed to be infinite-dimensional. Let $\H_V=\bigotimes_{x\in V}
\H_x$, for any finite $V\subset\Ir^d$, be the Hilbert space associated
with $V$. The unperturbed model has finite-volume Hamiltonians
of the form
\be
H^0_V=\sum_{x, V_0 + x\subset V} h_x\, ,
\ee
where $h_x$ is a selfadjoint operator acting non-trivially only on
$\H_{V_0 + x}$, for some finite $V_0$.
The main assumption is then that there exists
$0\neq\Omega_x\in\H_x$ such that 
$\Omega^0_V=\bigotimes_{x\in V}\Omega_x$ is the unique
zero-energy ground state of $H^0_V$, with  a
spectral gap of magnitude at least $\vert V_0\vert$ above the
ground state. Explicitly:
\be
H^0_V \geq 0\, , \quad H^0_V\Omega^0_V=0\, ,\quad
H_V\ge 
\vert V_0\vert (\idty-\vert\Omega^0_V\rangle\langle\Omega^0_V\vert)\, .
\label{a0}\ee

The perturbed Hamiltonians are assumed to be of the form
\be
H_V = H^0_V + \sum_{x,V_0 + x \subset V}\phi_x^{(r)} + \phi_x^{(b)},
\label{a1}\ee
where $\phi_x^{(r)}$ and $\phi_x^{(b)}$ are selfadjoint operators
on $\H_{V_0 +x}$ satisfying
\be
\vert\langle \psi, \phi_x^{(r)}\psi\rangle|\le\alpha\|h_x^{1/2}\psi\|^2\, ,
\quad 
\|\phi^{(b)}_x\|\le \beta\, ,
\label{a2}\ee
for all $\psi \in \dom (h_x^{1/2})$, and suitable constants $\alpha$ 
and $\beta$.
One can call $\phi^{(r)}$ a ``purely relatively bounded'' perturbation, while
$\phi^{(b)}$ is simply a bounded perturbation.

\begin{theorem}[Yarotsky \cite{yarotsky2004}]\label{thm:perturb}
Let $H_V$ of the form \eq{a1}, satisfying assumption \eq{a0}.
For all $\kappa>1$ there exists $\delta=\delta(\kappa,d,V_0)>0$ 
such that if condition \eq{a2} is
satisfied with some $\alpha\in (0,1)$, and 
$\beta=\delta(1-\alpha)^{\kappa(d+1)}$, then\newline
1) $H_V$ has a non-degenerate gapped ground state
$\Omega_V:$ $H_V\Omega_V=E_V\Omega_V,$ 
and for some $\gamma>0$, independent of $V$, we have
\be
H_V\ge E_V \vert\Omega_V\rangle\langle\Omega_V\vert
+(E_V + \gamma)(\idty
-\vert\Omega_V\rangle\langle\Omega_V\vert)\, .
\ee
2) There exists a thermodynamic weak$^*$-limit of the ground
states $\Omega_V:$ 
\be
\langle
A\Omega_V,\Omega_V\rangle
\xrightarrow{V\nearrow\Ir^d} \omega(A),
\ee 
where $A$ is a bounded local observable.
\newline
3) There is an exponential decay of correlations in the infinite
volume ground state $\omega$: for some positive $c$ and $\mu$,
and $A_i\in\B(\H_{V_i})$,
\be
|\omega(A_1A_2)-\omega(A_1)\omega(A_2)|\le
c^{|V_1|+|V_2|} e^{-\mu d(V_1,V_2)}\|A_1\|\|A_2\|\, .
\ee
4) If, within the allowed range of perturbations, the terms $\phi_x$
(or the resolvents $(h_x+\phi_x-z)^{-1}$ in the case of unbounded
perturbations) depend analytically on some parameters, then the
ground state $\omega$ is also weak$^*$ analytic in these
parameters (i.e. for any local observable $A$ its expectation
$\omega(A)$ is analytic).
\end{theorem}

Application of this result to the AKLT model \eq{aklt} yields the following
theorem.

\begin{theorem}\label{thm:perturb_aklt}
Let $\Phi=\Phi^*\in\A_{[0,r]}$. Then there exists $\lambda_0>0$, such that 
for all $\lambda, |\lambda|<\lambda_0$, the spin chain with Hamiltonian
$$
H = H^{\rm AKLT} + \lambda \sum_{x} \Phi_x
$$
has a unique infinite-volume ground state with a spectral gap and
exponential decay of correlations.
Here $\Phi_x\in A_{[x,x+r]}$, is $\Phi$ translated by $x$.
\end{theorem}
 
To prove this theorem, Yarotsky shows that the AKLT model itself
can be regarded as a perturbation of a particular model, one he
explicitly constructs, to which Theorem \ref{thm:perturb} can be applied.
 
\section{Ferromagnetic Ordering of Energy Levels}\label{sec:foel}

One easily checks that the Heisenberg Hamiltonian $H_V$, defined in
\eq{hh}, commutes with both the total spin matrices and the 
Casimir operator given by
$$
S^i_V=\sum_{x\in V} S^i_x, \, i = 1, 2, 3, \quad \text{and} \quad
C=\vec{S}_V \cdot\vec{S}_V.
$$
The eigenvalues of $C$ are $S(S+1)$ where the parameter $S
\in \{ S_{\rm min}, S_{\rm min}+1,\ldots,S_{\rm max} \}$, with
$S_{\rm max}=\sum_x s_x$. $S$ 
is called the total spin, and it labels the irreducible 
representations of $SU(2)$. Let $\H^{(S)}$ be the eigenspace
corresponding to those vectors of total spin $S$. One can show that 
$\H^{(S)}$ is an invariant subspace for the Hamiltonian $H_V$ and
therefore, the number
$$
E(H_V,S) := \min \spec H_V \vert_{\H^{(S)}},
$$
is well-defined. 

Supported by partial results and some numerical calculations, we made the following conjecture
in \cite{nachtergaele2004}.

\begin{conjecture}[\cite{nachtergaele2004}]
All ferromagnetic Heisenberg models have the
{\em Ferromagnetic Ordering of Energy Levels} (FOEL) 
property, meaning
$$
E(H_V,S)< E(H_V,S^\prime), \text{ if } S^\prime < S.
$$
\end{conjecture}

\begin{figure}
\begin{center}
\resizebox{!}{13truecm}{\includegraphics{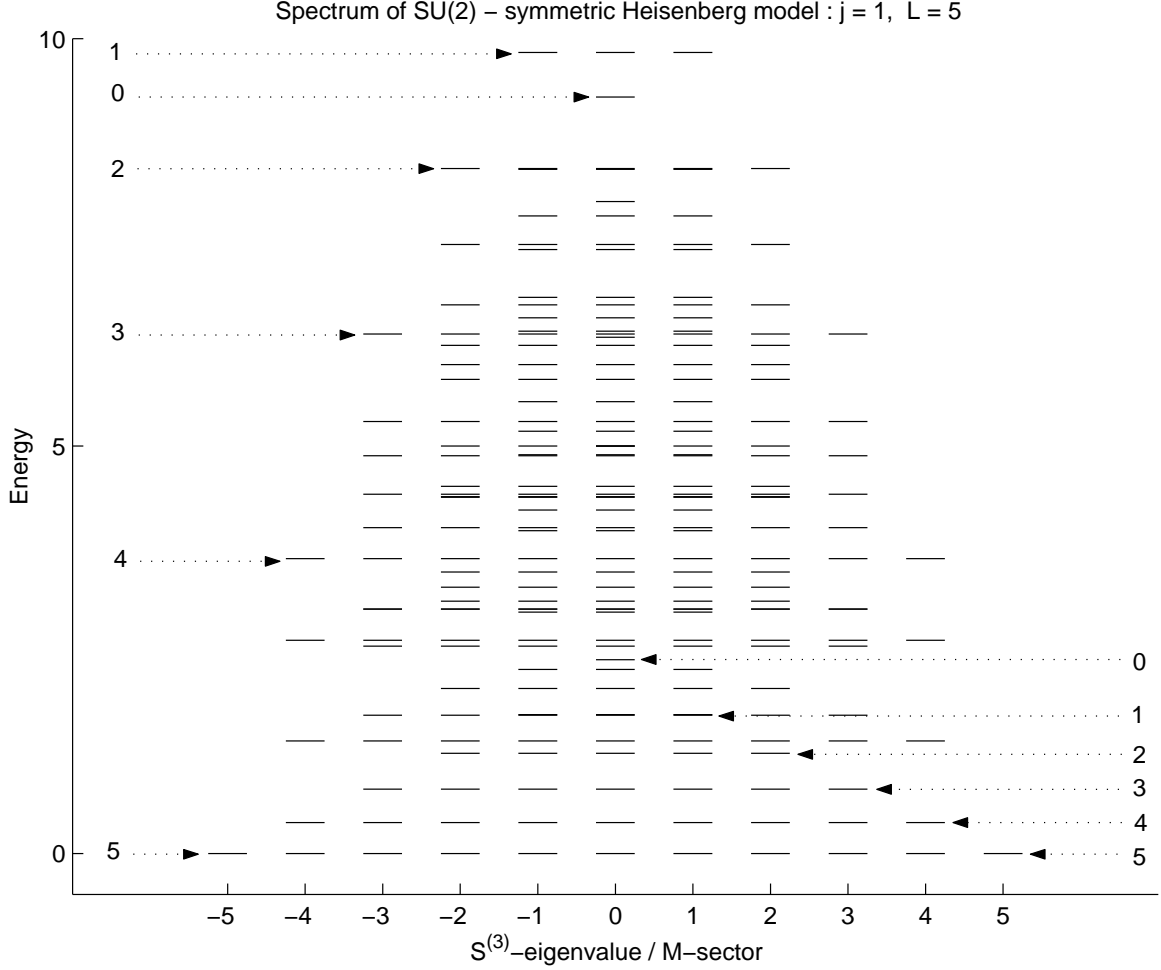}} 
\caption{\label{fig:foel}
The spectrum of a ferromagnetic Heisenberg chain  consisting of $5$ spin-$1$
spins with constant couplings.  On the horizontal axis we have plottted
the eigenvalue of the third component of the total spin. The spectrum is
off-set so that the ground state energy vanishes. The arrows on the right,
with  label $S$, indicate the multiplets of eigenvalues $E(H,S)$, i.e., the
smallest eigenvalue in the subspace of total spin $S$. The monotone ordering of
the spin labels is the FOEL property. On the left, we have indicated the
largest eigenvalues for each value of the total spin. The monotone ordering of
their labels in the  range $1,\ldots, 5$, is the content of the Lieb-Mattis
theorem \cite{lieb1962} applied to this system.}
\end{center}
\end{figure}

In \cite{lieb1962}, Lieb and Mattis proved ordering of energy levels 
for a class of Heisenberg models on bipartite graphs, which
includes the standard antiferromagnetic Heisenberg model. 
The FOEL property mentioned above can be considered as the 
ferromagnetic counterpart. To compare, a bipartite graph $G=(V,E)$ 
is a graph such that its set of vertices $V$ has a partition 
$V=A \cup B$ where $A \cap B = \emptyset$ and 
and any edge $(xy)\in E$ satisfies either $x\in A$ and $y\in B$, or 
$x \in B$ and $y \in A$. For such a graph, one considers 
Hamiltonians of the form
$$
H=-H_V+H_A + H_B,
$$ 
where $H_V,H_A$, and $H_B$ are ferromagnetic Heisenberg Hamiltonians
on the graph $G$, and arbirary graphs $A$ and $B$, respectively.
Let  $S_X=\sum_{x\in X} s_x$, for $X=A,B, V$.
The Lieb-Mattis Theorem \cite{lieb1962,lieb1989} then states that

(i) the ground state energy of $H$ is $E(H, \vert S_A - S_B\vert)$

(ii) if $\vert S_A - S_B\vert \leq S< S^\prime$, then $E(H, S) < E(H,S^\prime)$. 

One can see this property illustrated in Figure \ref{fig:foel}.

We first obtained a proof of FOEL for the spin 1/2 chain in 
\cite{nachtergaele2004}.
In that paper we also prove the same result for the ferromagnetic XXZ chain 
with $SU_q(2)$ symmetry. Later, in \cite{nachtergaele2005a}, we 
generalized the result to chains with arbitray values of the spin magnitudes
$s_x$ and coupling constants $J_{x,x+1}>0$. In short, we have the following
theorem.

\begin{theorem}
FOEL holds for all ferromagnetic chains.
\end{theorem}

The main tool in the proof is a special basis of SU(2) highest weight vectors introduced by
Temperley-Lieb \cite{Temperley1971} in the spin 1/2 case and by Frenkel and Khovanov 
\cite{frenkel1997} in the case of arbitrary spin. Some generalizations beyond 
the standard Heisenberg model have been announced  
\cite{nachtergaele2005a,nachtergaele2005b}.

The FOEL property has a number of interesting consequences.
The first immediate implication of FOEL is that the ground state energy of $H$
is $E(H,S_{\rm max})$, corresponding to the well-know fact that the ground state
space coincides with the subspace of maximal total spin.
Since there is only one multiplet of total spin $S_{\rm max}$, FOEL
also implies that the gap above the ground state is $E(H, S_{\rm max}-1)-E(H, S_{\rm max})$.
In the case of translation invariant models this is the physically expected property
 that the lowest excitations are simple spin-waves.

Another application of the FOEL property arises from the unitary equivalence of
the Heisenberg Hamiltonian and the generator of the Symmetric Simple Exclusion
Process (SSEP). To be precise, let $G=(V,E)$ be any finite graph, and define
$\Omega_n$  to be the configuration space of $n$ particles, for
$n=0,1,\ldots,\vert V\vert$, consisting of  $\eta:V\to \{0,1\}$, with 
$\sum_{x\in V} \eta(x)=n$. For any $(xy)\in E$, let $r_{xy}>0$. The SSEP is the
continuous time Markov process on $\Omega_n$ which exchanges the states
(whether there is a particle or not) at $x$ and $y$ with rate $r_{xy}$,
independently for each edge $(xy)$. The case $n=1$ is the random walk on $G$
with the given rates.

Alternatively, this process is defined by its generator on $l^2(\Omega_n)$:
$$
Lf(\eta) = \sum_{(xy)\in E}r_{xy}(f(\eta)-f(\eta^{xy})),
$$
where $\eta^{xy}$ is the configuration $\eta$ with the values at $x$ and $y$ 
interchanged. One verifies $L\geq 0$, $L1=0$, and therefore 
$L$ generates a Markov semigroup $\{e^{-tL}\}_{t\geq 0}$, such that
$$
\int f(\eta)\mu_t(d\eta)=\int (e^{-tL}f)(\eta) \mu_0(d\eta)\quad.
$$
where $\mu_0$ is the initial probability distribution on the particle 
configurations. It is easy to show that for each $n$ there is a unique
stationary measure given by the uniform distribution on $\Omega_n$. The
relaxation time, which determines the exponential rate of convergence to the
stationary state, is given by $1/\lambda(n)$, where $\lambda(n)>0$ is the
spectral gap (smallest eigenvalue $>0$) of $L$ as an operator on
$l^2(\Omega_n)$.

Aldous, based on discussions with Diaconis \cite{aldous}, made the following
remarkable conjecture concerning $\lambda(n)$:

\begin{conjecture}[Aldous]
$\lambda(n) = \lambda(1)$, for all $1\leq n\leq \vert V\vert -1$.
\end{conjecture}

Assuming the conjecture, one may determine the gap by solving the
one-particle problem.

To make the connection with FOEL, observe 
$$
\bigoplus_{n=0}^{\vert V\vert} l^2(\Omega_n) \cong (\Cx^2)^{\otimes\vert V \vert}\equiv \H_V
$$
This is the Hilbert space of a spin 1/2 model on $V$.
An explicit isomorphism is given by
$$
f\mapsto \psi=\sum_\eta f(\eta)\ket{\eta},
$$
where $\ket{\eta}\in\H$ is the tensor product basis vector defined by
$$
S^3_x\ket{\eta}=(\eta_x-1/2)\ket{\eta}
$$
Under this isomorphism the generator, $L$, of the SSEP becomes the
XXX Hamiltonian $H$. To see this it suffices to calculate the action
of $L$:
\beann
Lf(\eta)&\mapsto &\sum_\eta (Lf)(\eta)\ket{\eta}\\
&=&\sum_{(xy)}\sum_\eta r_{xy}(f(\eta)-f(\eta^{xy}))\ket{\eta}\\
&=&\sum_{(xy)} \sum_\eta r_{xy}(1-t_{xy}) f(\eta) \ket{\eta}\\
&=&H\psi
\eeann
where $t_{xy}$ is the unitary operator that interchanges 
the states at $x$ and $y$, and the last step uses
$$
1/2-2\vec{S}_x\cdot\vec{S}_y=1-t_{xy},
$$
and $J_{xy}=2r_{xy}$.

The number of particles, $n$, is a conserved quantity for SSEP. Since,
$S^3_{\rm tot} = -\vert V \vert/2 +n$, the corresponding conserved quantity for
the Heisenberg model is the third component of the total spin. Under the
isomorphism the invariant subspace of all functions $f$  supported on
$n$-particle configurations is identified with the set of vectors with
$S^3=S^3_{\rm max} -n$, which we will denote by $\H_n$. The unique invariant
measure of SSEP for $n$ particles on $V$,  which is the
uniform measure, corresponds to the ferromagnetic ground state with
magnetization $n -\vert V \vert/2$. The eigenvalue $\lambda(n)$ is the
spectral gap of $H_V\vert_{\H_n}$. It is then easy to see that the
FOEL property implies that $\lambda(n)=\lambda(1)$.  Since we proved FOEL
for chains, we also provided a new proof of Aldous' Conjecture in that case
\cite{handjani1996}. 

\section{Droplet Excitations of the XXZ Chain}
\label{sec:droplets}

The low-lying excitations of the ferromagnetic XXZ chain describe 
droplets, i.e., domains of opposite magnetization \cite{nachtergaele2001a}.
This is to be contrasted with the situation for the antiferromagnetic
XXZ chain. For the antiferromagnetic chain it has been proven that the 
low-energy spectrum does not contain droplet states;  states 
with an antiferromagnetically ordered domain which is out of phase with its 
surroundings \cite{datta2003,matsui2005}. In the following we only
discuss the ferromagnetic chain.

Kennedy calculated the droplet energies as a function of the quasi-momentum.
He obtained a Fourier series with coefficients that are a power series
in $1/\Delta$ \cite{kennedy2005}.  

By combining the Bethe Ansatz with the representation of the XXZ
Hamiltonian in the Temperley-Lieb basis employed to prove FOEL,
it has been possible to obtain more detailed information about the 
location and width of the energy band comprised of the droplet states
\cite{nachtergaele2005e}.

To state the new results we first need to introduce the 
$SU_q(2)$-symmetric XXZ chain \cite{pasquier1990}. 
Consider the spin 1/2 chain with boundary fields as defined by
the following Hamiltonian:
\beann
H_L&=&-\sum_{x=1}^{L-1} J [\Delta^{-1} (S^1_x S^1_{x+1} + 
S^2_x S^2_{x+1})+ (S_x^{3} S_{x+1}^{3} - 1/4)]\\
&&-A(\Delta)( S_L^3-S^3_1).
\eeann
where, $J>0$, $\Delta >1$, and $A(\Delta)=\frac{1}{2}\sqrt{1-1/\Delta^2}$.
This model commutes with $SU_q(2)$, with $q\in (0,1)$ such that 
$\Delta=(q+q^{-1})/2$. The generators of this symmetry are:
\beann
S^3\!&=&\!\sum_{x=1}^L \idty_1\otimes\cdots\otimes
S^3_x\otimes\idty_{x+1}\otimes\cdots\idty_L\label{spinmua}\\
S^+\!&=&\!\sum_{x=1}^L t_1\otimes\cdots\otimes t_{x-1}\otimes
S^+_x\otimes\idty_{x+1}\otimes\cdots\idty_L\label{spinmub}\\
S^-\!&=&\!\sum_{x=1}^L \idty_1\otimes\cdots\otimes
S^-_x\otimes t^{-1}_{x+1}\otimes\cdots t^{-1}_L
\eeann
where 
$$
t_x=\left(\begin{array}{cc} q^{-1}&0\\0&q\end{array}\right)\, , \quad 
[S^+,S^-]=\frac{q^{2S^3}-q^{-2S^3}}{q-q^{-1}} \, .
$$
The Hamiltonian $H_L$ also commutes with the Casimir opeator for $SU_q(2)$, 
given by 
$$
C= S^+S^- + \frac{ (qT)^{-1}+qT}{(q^{-1}-q)^2}, 
\quad T= t_1\otimes t_2\otimes\cdots\otimes t_L.
$$
The eigenvalues of $C$ are 
$$
 \frac{q^{-(2S+1)}+q^{2S+1}}{(q^{-1}-q)^2}, \quad S=0,1/2,1,3/2,\ldots
$$
and play the same role as $S$ for the XXX model, e.g., they label 
the irreducible representations of $SU_q(2)$.

The FOEL property with respect to $S$,  as defined in Section \ref{sec:foel}, 
can be proved in the same way as before \cite{nachtergaele2004}:
$$
E(H_L, S+1) <  E(H_L,S).
$$
It is rather natural to ask what the states with minimal energy for given $S$ 
describe. It will be convenient to express $S$ by its deviation from the
maximum possible value, i.e., $n$ such that $S=S_{\rm max} -n$. The answer is
that $E(H_L, S_{\rm max} -n)$, is the ground state energy of a droplet of $n$
down spins in a background of up spins. By the quantum group symmetry this is
necessarily also the energy of a state in which a kink and a droplet coexist.
In the thermodynamic limit the quantum symmetry evaporates, at least in the
ground state representation \cite{fannes1996}, but some traces remain. 
In particular, the droplet energies computed in the $SU_q(2)$-symmetric
model give the correct droplet energy in thermodynamic limit computed with
periodic boundary conditions. To state this precisely,  define $E_L(n)=\inf
\spec H^{\rm periodic}_L\vert_{\H_n}$, where $\H_n$ is the eigenspace of $S^3$
with eigenvalue $S^3_{\rm max}-n$. The following theorem is proved in 
\cite{nachtergaele2005e}:

\begin{theorem}
$$
\lim_{L\to\infty} E(H_L,S_{\rm max} -n)
=\lim_{L\to\infty} E_L(n)=E(n)\equiv\frac{(1-q^2)(1-q^n)}{(1+q^2)(1+q^n)}.
$$
\end{theorem}

The value of $E(n)$ itself is calculated by a simple version of the Bethe
Ansatz. To make the calculation rigorous, we rely on positivity properties of
the Hamiltonian in suitable bases.  One can further show that $E(n)$ belongs to
the continuous spectrum and is the bottom of a band of width
$$
4q^n\frac{1-q^2}{1-q^{2n}}.
$$
The states corresponding to this band can be interpreted as a droplet of size
$n$  with a definite momentum. If we consider the droplet as a particle,
the formula for the width indicates that the ``mass'' of the particle 
diverges as $n\to \infty$.

The energy $E(n)$ was considered by Yang and Yang in their famous series of
papers on the XXZ chain \cite{yang1966c}. However, due to an error, they got an
energy  of order $n$, which prevented the interpretation of the states as
droplet states.

\bigskip
\noindent
{\em Acknowledgement.} This work was supported in part by the
National Science Foundation under Grant \# DMS-0303316.


\end{document}